\documentclass[onecolumn, 11pt, draftclsnofoot, letterpaper]{IEEEtran}
\ifCLASSINFOpdf
  \usepackage[dvips]{graphicx}
 \else
 	\usepackage[dvips]{graphicx}
  \fi
\usepackage{amsmath, amsfonts, amssymb, xcolor, lipsum}
\usepackage{mleftright}

\DeclareMathOperator*{\argmin}{arg\,min}

\begin{document}
\title{Transmit Antenna Selection for Physical-Layer Network Coding Based on Euclidean Distance}

\author{\IEEEauthorblockN{Vaibhav Kumar\IEEEauthorrefmark{1},
Barry Cardiff\IEEEauthorrefmark{2}, and Mark F. Flanagan\IEEEauthorrefmark{3}} \\
\IEEEauthorblockA{School of Electrical and Electronic Engineering, \\
University College Dublin, Belfield, Dublin 4, Ireland\\
Email: \IEEEauthorrefmark{1}vaibhav.kumar@ucdconnect.ie,
\IEEEauthorrefmark{2}barry.cardiff@ucd.ie,
\IEEEauthorrefmark{3}mark.flanagan@ieee.org}}

% after \documentclass[•]{•}

%Telephone: (800) 555--1212, Fax: (888) 555--1212}


\maketitle

\begin{abstract}
Physical-layer network coding (PNC) is now well-known as a potential candidate for delay-sensitive and spectrally efficient communication applications, especially in two-way relay channels (TWRCs). In this paper, we present the error performance analysis of a multiple-input single-output (MISO) fixed network coding (FNC) system with two different transmit antenna selection (TAS) schemes. For the first scheme, where the antenna selection is performed based on the strongest channel, we derive a tight closed-form upper bound on the average symbol error rate (SER) with $M$-ary modulation and show that the system achieves a diversity order of 1 for $M > 2$. Next, we propose a Euclidean distance (ED) based antenna selection scheme which outperforms the first scheme in terms of error performance and is shown to achieve a diversity order lower bounded by the minimum of the number of antennas at the two users.
\end{abstract}


\IEEEpeerreviewmaketitle

\section{Introduction}
Wireless PNC has received a lot of attention among researchers in recent years due to its inherent desirable properties of delay reduction, throughput enhancement and better spectral efficiency. The advantage of PNC can easily be seen in a TWRC, where bidirectional information exchange takes place in the half-duplex mode between two users $A$ and $B$ with the help of a relay $R$. In a TWRC, PNC requires only two time slots to exchange the information between the users compared to three time slots required by traditional network coding \cite{Liew}. In the first time slot, also termed the \emph{multiple access} (MA) phase, both users $A$ and $B$ simultaneously transmit their data to the relay $R$. Based on its received signal, the relay forms the maximum-likelihood (ML) estimate of the \emph{pair} of transmitted user constellation symbols. This estimate of the pair of user symbols is then mapped to a \emph{network-coded} constellation symbol using the denoise-and-forward (DNF) protocol \cite{Popovski} and the relay broadcasts this to both users in the next time slot, called the \emph{broadcast} (BC) phase. User constellation symbol pairs which are mapped to the same complex number in the network-coded constellation are said to form a \emph{cluster}. Using its own message transmitted in the previous MA phase, $A$ can decode the message transmitted from $B$ and vice versa. 

In the case of a \emph{fixed network coding} (FNC) system, the network code applied at the relay is always fixed and does not depend on channel conditions. One of the bottlenecks in the FNC system limiting the error performance is the existence of \emph{singular fade states} \cite{RajanAdaptive} that result in shortening the distance between clusters and making the relay vulnerable to erroneous mapping. In order to mitigate the distance shortening phenomenon, adaptive network coding (ANC) has been proposed in \cite{Akino}, where the network coding applied at the relay varies with the channel conditions. The network coding scheme at the relay in ANC systems depends on the ratio of the channel coefficients between the $A$-$R$ and $B$-$R$ links. For QPSK modulation in the MA phase, an unconventional 5-ary modulation scheme has been suggested in \cite{Akino} for the BC phase and a computer search algorithm called \emph{closest-neighbor clustering} (CNC) has been proposed to obtain adaptive codes. An analytical treatment for the ANC scheme has been presented in \cite{RajanAdaptive} considering a Rician fading model. Although the ANC scheme alleviates the problem of distance shortening between the clusters in an efficient way, the related system complexity increases significantly. 

In \cite{Huang}, the error performance of a multiple-antenna based PNC system has been investigated in detail considering a Rayleigh fading model and BPSK modulation. A transmit antenna selection (TAS) scheme based on the strongest channel (in terms of signal-to-noise ratio) between the user and the relay has been applied in \cite{Huang} where the relay, which is assumed to have perfect channel state information (CSI), decides on the indices of the antenna to be used at $A$ and $B$, and shares this information with the users via a error-free, low bandwidth feedback channel. It has been shown in \cite{Huang} that for the case  when both users $A$ and $B$ are equipped with multiple antennas and the relay has only one antenna (MISO), the diversity order is equal to the minimum of the number of antennas at the two users. We show analytically that this result is true only for the case of binary modulation, while for higher-order modulation the antenna selection scheme based on the strongest channel fails to exploit the advantage of multiple antennas at the user end to leverage diversity gain.  

The contribution made in the present paper is twofold. First, we analyze the error performance of the system presented in \cite{Huang} with $M$-ary modulation for the MISO case and present a closed-form expression for a tight upper bound on the average SER at the end of the MA phase. Second, we propose a new TAS scheme that maximizes the minimum ED between different clusters at the relay for the fixed network coded PNC system. To the best of our knowledge, the analysis of a PNC system with a ED based antenna selection scheme is not yet available in the open literature. We also prove analytically that the diversity order of such a PNC system is lower bounded by the minimum of the number of antennas at the user end. Since our system uses a fixed network coding scheme, the implementation complexity is lower than that involved in ANC systems, and we do not require the use of any nonstandard (e.g. 5-ary) modulation scheme in the BC phase, reducing the complexity associated with the constellation design. 

\section{System Model}
\begin{figure}[hbht]
\centering
\includegraphics[scale=01]{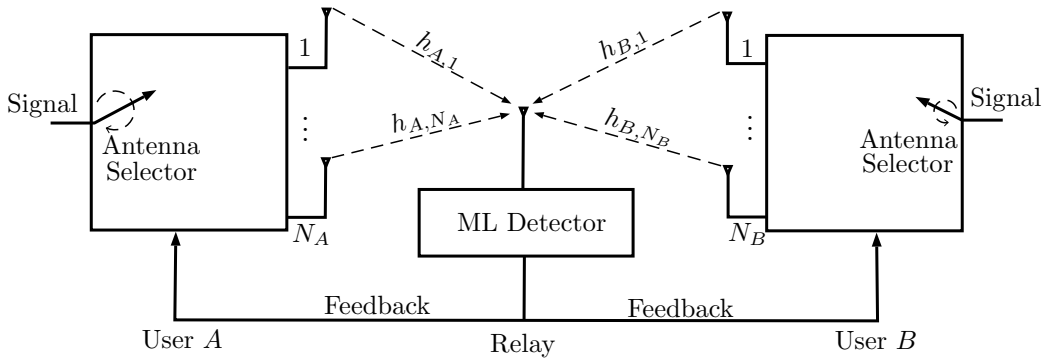}
\caption{System model for PNC with multiple transmit antennas and a single receive antenna.}
\end{figure}
The system model is shown in Fig. 1, where the users $A$ and $B$ are equipped with $N_{A}$ ($>1$) and $N_{B}$ ($>1$) antennas respectively, while the relay $R$ has only a single antenna. During every transmission, only one of the antennas from each user is used for transmission, and the choice of antennas is based on feedback received from the relay. The channel between user $m \in \{A, B\}$ and the relay $R$ is modeled as slow Rayleigh fading with perfect CSI available at $R$ only. We assume that the channel remains constant during a frame transmission and changes independently from one frame to another. Hence the channel coefficient between the $i^{\text{th}}$ antenna of user $m$ and the relay is distributed according to $\mathcal{CN}(0, 1)$. Suppose that both users employ the same unit-energy $M$-ary constellation $\mathcal{S}$ and $\Delta S$ denotes the difference constellation set of $\mathcal{S}$, defined as $\Delta \mathcal{S} = \{x - x^{\prime}|x, x^{\prime} \in \mathcal{S}\}$. Letting $s_{m} \in \mathbb{Z}_{M} = \{0, 1, ,\ldots, M-1\}$ denote the message symbol at user $m$, and $x_{m} \in \mathcal{S}$ denote the corresponding transmitted constellation symbol, the signal received at the relay during the MA phase is
\begin{equation}
	y = \sqrt{E_{s}}h_{A} x_{A} + \sqrt{E_{s}} h_{B} x_{B} + n
\end{equation}
where $n$ denotes the noise at the relay and is assumed to be distributed according to $\mathcal{CN}(0, N_{0})$, $E_{s}$ denotes the energy of the transmitted signal and $h_{m}$ is the channel coefficient of the link between the selected antenna of user $m$ and the relay (based on the antenna selection scheme). The ML estimate of the transmitted symbol pair $(x_{A}, x_{B}) \in \mathcal{S}^{2}$ is given by
\begin{equation}
	\left(\hat{x}_{A}, \hat{x}_{B}\right) = \argmin_{(x_{A}, x_{B}) \in \mathcal{S}^{2}} \left\vert y - \sqrt{E_{s}}h_{A}x_{A} - \sqrt{E_{s}}h_{B}x_{B} \right\vert.
\end{equation}

The relay then maps this estimate $(\hat{x}_{A}, \hat{x}_{B}) \in \mathcal{S}^{2}$ to the network coded symbol $\hat{x}_{R} \in \mathcal{S}$ using the PNC map $\mathcal{M}: \mathcal{S}^{2} \rightarrow \mathcal{S}$. Using their own message $x_{m}$, transmitted in the previous MA phase, each user can decode the other user's message, provided that the map $\mathcal{M}$ satisfies the \emph{exclusive law} \cite{Akino}. Table~I shows the network coding operation for QPSK modulation, where $s_{R} = s_{A} \oplus s_{B} \in \mathbb{Z}_{4}$ and $\oplus$ denotes \emph{bitwise} addition (XOR) in $\mathbb{Z}_{4}$.
\begin{table}[hbht]
\centering
\caption{Example PNC mapping at the relay for QPSK constellation.}
\label{my-label}
\begin{tabular}{|c|c|c|c|}
\hline
$(s_{A}, s_{B})$                                                        & $(x_{A}, x_{B})$                                                                                    & $s_{R}$ & \begin{tabular}[c]{@{}c@{}} $x_{R} = $\\ {\tiny{$\mathcal{M}(x_{A}, x_{B})$}}\end{tabular} \\ \hline
\begin{tabular}[c]{@{}c@{}}(0, 0), (1, 1),\\ (2, 2), (3, 3)\end{tabular} & \begin{tabular}[c]{@{}c@{}}$\left(\frac{1+i}{\sqrt{2}}, \frac{1+i}{\sqrt{2}} \right)$, $\left(\frac{-1+i}{\sqrt{2}}, \frac{-1+i}{\sqrt{2}}\right)$,\\ $\left(\frac{-1-i}{\sqrt{2}}, \frac{-1-i}{\sqrt{2}}\right)$, $\left(\frac{1-i}{\sqrt{2}}, \frac{1-i}{\sqrt{2}}\right)$\end{tabular} & 0       & $\frac{1+i}{\sqrt{2}}$     \\ \hline
\begin{tabular}[c]{@{}c@{}}(0, 1), (1, 0),\\ (2, 3), (3, 2)\end{tabular} & \begin{tabular}[c]{@{}c@{}}$\left(\frac{1+i}{\sqrt{2}}, \frac{-1+i}{\sqrt{2}}\right)$, $\left(\frac{-1+i}{\sqrt{2}}, \frac{1+i}{\sqrt{2}}\right)$,\\ $\left(\frac{-1-i}{\sqrt{2}}, \frac{1-i}{\sqrt{2}}\right)$, $\left(\frac{1-i}{\sqrt{2}}, \frac{-1-i}{\sqrt{2}}\right)$\end{tabular} & 1       & $\frac{-1+i}{\sqrt{2}}$    \\ \hline
\begin{tabular}[c]{@{}c@{}}(0, 2), (2,0),\\ (1, 3), (3,1)\end{tabular}   & \begin{tabular}[c]{@{}c@{}}$\left(\frac{1+i}{\sqrt{2}}, \frac{-1-i}{\sqrt{2}}\right)$, $\left(\frac{-1-i}{\sqrt{2}}, \frac{1+i}{\sqrt{2}}\right)$,\\ $\left(\frac{-1+i}{\sqrt{2}}, \frac{1-i}{\sqrt{2}}\right)$, $\left(\frac{1-i}{\sqrt{2}}, \frac{-1+i}{\sqrt{2}}\right)$\end{tabular} & 2       & $\frac{-1-i}{\sqrt{2}}$    \\ \hline
\begin{tabular}[c]{@{}c@{}}(0, 3), (3,0),\\ (1, 2), (2, 1)\end{tabular}  & \begin{tabular}[c]{@{}c@{}}$\left(\frac{1+i}{\sqrt{2}}, \frac{1-i}{\sqrt{2}}\right)$, $\left(\frac{1-i}{\sqrt{2}}, \frac{1+i}{\sqrt{2}}\right)$,\\ $\left(\frac{-1+i}{\sqrt{2}}, \frac{-1-i}{\sqrt{2}}\right)$, $\left(\frac{-1-i}{\sqrt{2}}, \frac{-1+i}{\sqrt{2}}\right)$\end{tabular} & 3       & $\frac{1-i}{\sqrt{2}}$     \\ \hline
\end{tabular}
\end{table}

The performance analysis of the two TAS schemes is presented in detail in the next section. 
\section{Transmit Antenna Selection}
This section presents the analysis of two different TAS schemes for the PNC system. In the first scheme (TAS1), the index of the selected antenna at each user is the one having the highest signal-to-noise ratio (SNR). The channel coefficient between the selected antenna of user $m$ and the relay $R$ is given by
\begin{equation}
	h_{m} = \arg \max_{1 \leq i \leq N_{m}} \vert h_{m, i} \vert^{2}
\end{equation}
where $m \in \{A, B\}$ and $h_{m, i}$ is the channel coefficient between the $i^{\text{th}}$ antenna of user $m$ and relay $R$. 

In contrast to this, in TAS2 the transmit antenna of each user is selected such that the minimum ED between the clusters at the relay is maximized. A similar scheme has been discussed in \cite{Hari} for spatial modulation (SM). Let $\mathcal{I} = \{(i, j): 1 \leq i \leq N_{A}, 1 \leq j \leq N_{B}\}$ be the set which enumerates all of the possible $n = N_{A} \times N_{B}$ combinations of selecting one antenna from each user. Among these $n$ combinations, the set of transmit antennas that maximizes the minimum ED between the clusters is obtained as \cite[eqn.~(2)]{Hari}
\begin{equation}
	I_{ED} = \arg \max_{I \in \mathcal{I}} \left\{ \min_{\substack{\boldsymbol{x}, \boldsymbol{x}^{\prime} \in \mathcal{S}^{2} \\ \mathcal{M}(\boldsymbol{x}) \neq \mathcal{M}(\boldsymbol{x}^{\prime})}} \left\Vert \boldsymbol{H}_{I}(\boldsymbol{x} - \boldsymbol{x}^{\prime})\right\Vert^{2}\right\}
\end{equation}
where $\boldsymbol{H}_{I} = [h_{A, i} \ h_{B, j}] \in \mathbb{C}^{1 \times 2}$, $\boldsymbol{x} = [x_{A} \ x_{B}]^{T} \in \mathbb{C}^{2 \times 1}$, $\boldsymbol{x}^{\prime} = [x_{A}^{\prime} \ x_{B}^{\prime}]^{T} \in \mathbb{C}^{2 \times 1}$ and $\boldsymbol{H}_{I_{ED}} = [h_{A} \ h_{B}] \in \mathbb{C}^{1 \times 2}$ is the optimal channel vector. 

To understand the performance superiority of TAS2 over TAS1, we first consider a simple example of one transmission slot where the users transmit their messages using QPSK modulation. Suppose that $N_{A} = N_{B} = 2$ and $h_{A,1} = (1+i)/\sqrt{2}$, $h_{A,2} = (1 - 0.5i)/\sqrt{2}$, $h_{B,1} = (1 - 0.8i)/\sqrt{2}$ and $h_{B,2} = (1+0.7i)/\sqrt{2}$. In this case, since the $A_{1}$-$R$ link is stronger (i.e., has a higher SNR) than the $A_{2}$-$R$ link, and similarly the $B_{1}$-$R$ link is stronger than the $B_{2}$-$R$ link, TAS1 will choose the antenna combination $(A_{1}, B_{1})$. With this combination the minimum distance between the clusters at the relay becomes very small, which can lead to an incorrect ML estimate at the relay. Fig. 2 shows a plot, for TAS1, of the noise-free received signal at the relay, i.e., $h_{A}x_{A} + h_{B}x_{B}$ (here we assume $E_{s} = 1$), together with the corresponding network coded symbols, where each $2$-tuple in the figure represents $(s_{A}, s_{B})$. In contrast to this, the proposed antenna selection scheme (TAS2) chooses $(A_{1}, B_{2})$ as the optimal combination and the resulting network coded symbols are shown in Fig. 3. It is clear that TAS2 overcomes the distance shortening phenomenon.
\begin{figure}[hbht]
\centering 
\includegraphics[scale = 0.5]{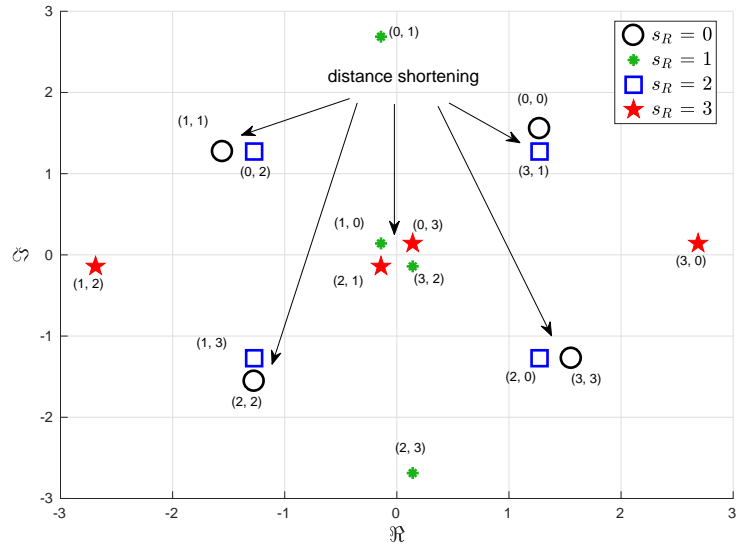}
\caption{Network coded symbols at the relay for QPSK constellation using TAS1.}
\end{figure}  
\begin{figure}[hbht]
\centering 
\includegraphics[scale = 0.7]{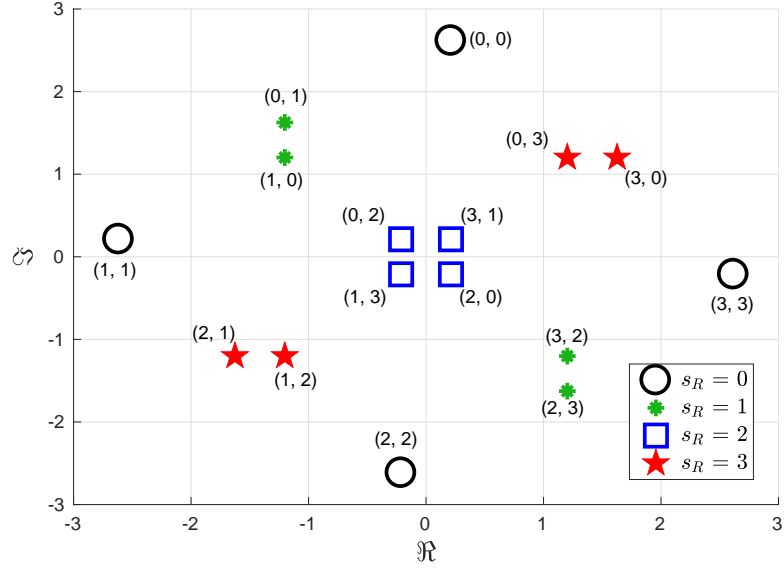}
\caption{Network coded symbols at the relay for QPSK constellation using TAS2.}
\end{figure}

In the following subsections, we present the analysis of the error performance for the two TAS schemes.
\subsection{TAS1: Antenna selection based on the maximum SNR}
For the error performance analysis of TAS1, we use the \emph{union-bound} approach given in \cite{RajanAdaptive} rather that the approach given in \cite{Huang} which applies only for binary modulations. For FNC, the average SER at the end of the MA phase is given by \cite[eqn.~(6)]{RajanAdaptive}
\begin{align}
	P_{e} & = \dfrac{1}{M^{2}} \sum_{(x_{A}, x_{B}) \in \mathcal{S}^{2}} \sum_{\substack{(x_{A}, x_{B}) \neq (x_{A}^{\prime}, x_{B}^{\prime}) \in \mathcal{S}^{2} \\ \mathcal{M}(x_{A}, x_{B}) \neq  \mathcal{M}(x_{A}^{\prime}, x_{B}^{\prime})}}  \mathbb{E} \left[P\left\{(\hat{x}_{A}, \hat{x}_{B}) = (x_{A}^{\prime}, x_{B}^{\prime}) | h_{A}, h_{B}\right\} \right] \nonumber \\
 & = \dfrac{1}{M^{2}} \left[ \sum_{(x_{A}, x_{B}) \in \mathcal{S}^{2}} \sum_{x_{A} \neq x_{A}^{\prime} \in \mathcal{S}} \mathbb{E} \left[P \{ (\hat{x}_{A}, \hat{x}_{B}) = (x_{A}^{\prime}, x_{B}) |h_{A}, h_{B} \}] \right. \right. \nonumber \\
	&  \hspace{1cm}+ \sum_{(x_{A}, x_{B}) \in \mathcal{S}^{2}} \sum_{x_{B} \neq x_{B}^{\prime} \in \mathcal{S}} \mathbb{E} \left[P \{ (\hat{x}_{A}, \hat{x}_{B}) = (x_{A}, x_{B}^{\prime}) |h_{A}, h_{B} \}\right]   \nonumber \\
	& \hspace{1cm} + \sum_{(x_{A}, x_{B}) \in \mathcal{S}^{2}} \sum_{\substack{ x_{A} \neq x_{A}^{\prime} \in \mathcal{S} \\ x_{B} \neq x_{B}^{\prime} \in \mathcal{S}}} \mathbb{E} [P \{ \hat{x}_{A}, \hat{x}_{B}) = (x_{A}^{\prime}, x_{B}^{\prime}), \mathcal{M}(x_{A}, x_{B}) \neq \mathcal{M}(x_{A}^{\prime}, x_{B}^{\prime}) | h_{A}, h_{B}\}\Bigg]
\end{align}
where $M$ is the modulation order and $\mathbb{E}(\cdot)$ is the expectation operator. An upper bound on the average SER can be given as
\begin{align}	
	P_{e} \leq \dfrac{1}{M^{2}} & \sum_{(x_{A}, x_{B}) \in \mathcal{S}^{2}}\sum_{ \substack{ (x_{A}, x_{B}) \neq (x_{A}^{\prime}, x_{B}^{\prime}) \in \mathcal{S}^{2} \\ \mathcal{M}(x_{A}, x_{B}) \neq \mathcal{M} (x_{A}^{\prime}, x_{B}^{\prime })} }\mathbb{E} \left[ Q \left( \sqrt{\dfrac{E_{s}}{2N_{0}}} \left\vert (h_{A}x_{A} + h_{B} x_{B}) - (h_{A}x_{A}^{\prime} + h_{B} x_{B}^{\prime}) \right\vert\right) \right]
\end{align}
where $Q(\cdot)$ is the Gaussian Q-function. The Chernoff bound on the Q-function used in \cite{RajanAdaptive} results in a loose upper bound for the present case and hence we use the Chiani approximation \cite[eqn.~(14)]{Chiani} instead, yielding
\begin{align}
	P_{e} \lesssim & \dfrac{1}{M^{2}} \sum_{(x_{A}, x_{B}) \in \mathcal{S}^{2}} \sum_{\substack{(x_{A}, x_{B}) \neq (x_{A}^{\prime}, x_{B}^{\prime}) \in \mathcal{S}^{2} \\ \mathcal{M}(x_{A}, x_{B}) \neq  \mathcal{M}(x_{A}^{\prime}, x_{B}^{\prime})}} \left[ \dfrac{1}{12} \mathbb{E} \left\{ \exp \left(-\dfrac{E_{s}}{4N_{0}} \left\vert h_{A} \Delta x_{A} + h_{B} \Delta x_{B} \right\vert^{2} \right) \right\} \right. \nonumber \\
	&  \hspace{6cm} + \left. \dfrac{1}{4} \mathbb{E} \left\{ \exp \left(-\dfrac{E_{s}}{3N_{0}} \left\vert h_{A} \Delta x_{A} + h_{B} \Delta x_{B} \right\vert^{2} \right) \right\} \right] \nonumber \\
	 = & \dfrac{1}{M^{2}} \sum_{(x_{A}, x_{B}) \in \mathcal{S}^{2}} \sum_{\substack{(x_{A}, x_{B}) \neq (x_{A}^{\prime}, x_{B}^{\prime}) \in \mathcal{S}^{2} \\ \mathcal{M}(x_{A}^{\prime}, x_{B}^{\prime}) \neq  \mathcal{M}(x_{A}, x_{B})}}  \left[\dfrac{1}{12} \mathbb{E}(\Upsilon_{1}) + \dfrac{1}{4} \mathbb{E}(\Upsilon_{2})\right].
\end{align}

Now we analyze the three different terms on the right-hand side of (5) separately as follows: 

\subsection*{Case I: When $x_{A} \neq x_{A}^{\prime}$ and $x_{B} = x_{B}^{\prime}$}
 In this case $\Delta x_{B} = 0$ and $\Upsilon_{1} = \exp \left( -\frac{E_{s}}{4N_{0}} |h_{A} \Delta x_{A}|^{2}\right)$. Hence, 
 \begin{align}
 	\mathbb{E}(\Upsilon_{1}) = \int_{0}^{\infty}\exp \left(- \dfrac{E_{s}}{4N_{0}} |h_{A} \Delta x_{A}|^{2}\right) f(|h_{A}|) d|h_{A}|
 \end{align}
where $f(\cdot)$ denotes the probability density function. Using the fact that each $|h_{m, i}|,\  m \in \{A, B\}, \ 1 \leq i \leq N_{m}$ is Rayleigh distributed, the distribution of the magnitude of $h_{m}$ can easily be derived as \cite[eqn.~(37)]{Huang}
 \begin{equation}
	f(|h_{m}|) = \sum_{k = 1}^{N_{m}} \binom{N_{m}}{k}(-1)^{k-1}2k|h_{m}| \exp(-k|h_{m}|^{2}).
\end{equation}

Substituting $f(|h_{m}|)$ into (8) we can write
\begin{align}
	\mathbb{E}(\Upsilon_{1}) = & \sum_{k=1}^{N_{A}} \binom{N_{A}}{k} (-1)^{(k-1)} k \int_{0}^{\infty} 2|h_{A}|  \exp\left\{-\left( k + \dfrac{E_{s}|\Delta x_{A}|^{2}}{4N_{0}}\right) |h_{A}|^{2}\right\}  \ d|h_{A}| \nonumber \\
	= & \sum_{k=1}^{N_{A}} \binom{N_{A}}{k} (-1)^{(k-1)} \left(1+\dfrac{E_{s}|\Delta x_{A}|^{2}}{4kN_{0}}\right) ^{-1}.
\end{align}

Similarly, 
\begin{align}
	\mathbb{E}(\Upsilon_{2}) = \sum_{k=1}^{N_{A}} \binom{N_{A}}{k} (-1)^{(k-1)} \left(1+\dfrac{E_{s}|\Delta x_{A}|^{2}}{3kN_{0}}\right) ^{-1}.
\end{align}

Substituting the values of $\mathbb{E}(\Upsilon_{1})$ and $\mathbb{E}(\Upsilon_{2})$ from (10) and (11) respectively into (7), the average SER arising from the case when  $x_{A} \neq x_{A}^{\prime}$ and $x_{B} = x_{B}^{\prime}$ can be written as
\begin{align}
	 P_{e}\{(\hat{x}_{A}, \hat{x}_{B}) = (x_{A}^{\prime}, x_{B})\} & \lesssim \dfrac{1}{M^{2}} \sum_{(x_{A}, x_{B}) \in \mathcal{S}^{2}} \sum_{x_{A} \neq x_{A}^{\prime} \in \mathcal{S}}  \sum_{k=1}^{N_{A}} \binom{N_{A}}{k} (-1)^{(k-1)}  \left[(12\Psi_{A,k})^{-1} + (4\Xi_{A,k})^{-1} \right] \nonumber \\
	= & \ \dfrac{1}{M^{2}} \sum_{(x_{A}, x_{B}) \in \mathcal{S}^{2}} \sum_{x_{A} \neq x_{A}^{\prime} \in \mathcal{S}}(\zeta_{1} + \zeta_{2})
\end{align}
where
\begin{align}
	&\Psi_{A,k} = 1+\dfrac{E_{s}|\Delta x_{A}|^{2}}{4kN_{0}}, \qquad \Xi_{A,k} = 1+\dfrac{E_{s}|\Delta x_{A}|^{2}}{3kN_{0}}, \\
	&\zeta_{1} =  \sum_{k=1}^{N_{A}} \binom{N_{A}}{k} \dfrac{(-1)^{(k-1)}}{12\Psi_{A,k}}, \zeta_{2} =  \sum_{k=1}^{N_{A}} \binom{N_{A}}{k} \dfrac{(-1)^{(k-1)}}{4\Xi_{A,k}}.
\end{align}

Using the binomial expansion, $\zeta_{1}$ can be written as
\begin{align}
	& \zeta_{1} = \dfrac{1}{12}\sum_{k=1}^{N_{A}} \binom{N_{A}}{k} \dfrac{(-1)^{(k-1)}}{1+\dfrac{E_{s}|\Delta x_{A}|^{2}}{4kN_{0}}} \nonumber \\
			    = & \dfrac{1}{12}\sum_{k=1}^{N_{A}} \binom{N_{A}}{k} \dfrac{(-1)^{(k-1)}}{\dfrac{E_{s}|\Delta x_{A}|^{2}}{4kN_{0}}} \left(1+ \dfrac{1}{\dfrac{E_{s}|\Delta x_{A}|^{2}}{4kN_{0}}} \right)^{-1} \nonumber \\
			  = & \dfrac{1}{12}\sum_{k=1}^{N_{A}} \binom{N_{A}}{k} \dfrac{(-1)^{(k-1)}}{\dfrac{E_{s}|\Delta x_{A}|^{2}}{4kN_{0}}} \sum_{n = 1}^{\infty} \left(\dfrac{-1}{\dfrac{E_{s}|\Delta x_{A}|^{2}}{4kN_{0}}} \right)^{n-1} \nonumber \\
			  = & \sum_{n = 1}^{\infty} \left(\dfrac{E_{s}}{N_{0}}\right)^{-n} \underbrace{\sum_{k = 1}^{N_{A}} \binom{N_{A}}{k} \left(\dfrac{|\Delta x_{A}|^{2}}{4k}\right)^{-n}\dfrac{(-1)^{k+n-2}}{12}}_{C_{-n}} \nonumber \\
			 = & \sum_{n = N_{A}}^{\infty} \left(\dfrac{E_{s}}{N_{0}}\right)^{-n} C_{-n}\tag{$\because C_{-n} = 0 \ \forall \ 1 \leq n < N_{A}$} \\
			  < & \ \left(\dfrac{E_{s}}{N_{0}}\right)^{-N_{A}}C_{-N_{A}} + O \left[ \left( \dfrac{E_{s}}{N_{0}}\right)^{-N_{A}}\right]
\end{align}
where $O(\cdot)$ is the Landau symbol. Similarly, for $\zeta_{2}$, it can be shown that
\begin{equation}
	\zeta_{2} <  \left(\dfrac{E_{s}}{N_{0}}\right)^{-N_{A}} C_{-N_{A}}^{\prime} + O \left[ \left( \dfrac{E_{s}}{N_{0}}\right)^{-N_{A}}\right]
\end{equation}
where
\begin{equation}
	C_{-n}^{\prime} = \sum_{k = 1}^{N_{A}} \binom{N_{A}}{k} \left(\dfrac{|\Delta x_{A}|^{2}}{3k}\right)^{-n}\dfrac{(-1)^{k+n-2}}{4}.
\end{equation}

From (12), (15) and (16), it is clear that the average symbol error probability arising from the case when $x_{A} \neq x_{A}^{\prime}$ and $x_{B} = x_{B}^{\prime}$ decays as $(E_{s}/N_{0})^{-N_{A}}$ for higher values of $E_{s}/N_{0}$.
\subsection*{Case II: When $x_{A} = x_{A}^{\prime}$ and $x_{B} \neq x_{B}^{\prime}$}
In this case $\Delta x_{A} = 0$ and analogous to the previous case, the average symbol error probability can be given by
\begin{align}
	& P_{e}\{(\hat{x}_{A}, \hat{x}_{B}) = (x_{A}, x_{B}^{\prime})\} \lesssim \dfrac{1}{M^{2}} \sum_{(x_{A}, x_{B}) \in \mathcal{S}^{2}} \sum_{x_{B} \neq x_{B}^{\prime} \in \mathcal{S}} \nonumber \\
	& \sum_{k=1}^{N_{B}} \binom{N_{B}}{k} (-1)^{(k-1)}  \left[(12\Psi_{B,k})^{-1} + (4\Xi_{B,k})^{-1} \right]
\end{align}
where 
\begin{equation}
	\Psi_{B,k} = 1+\dfrac{E_{s}|\Delta x_{B}|^{2}}{4kN_{0}}, \qquad \Xi_{B,k} = 1+\dfrac{E_{s}|\Delta x_{B}|^{2}}{3kN_{0}}.
\end{equation}

Analogous to the previous case, the average error probability due to the case when $x_{A} = x_{A}^{\prime}$ and $x_{B} \neq x_{B}^{\prime}$ decays as $(E_{s}/N_{0})^{-N_{B}}$ for higher values of $E_{s}/N_{0}$.

\subsection*{Case III: When $x_{A} \neq x_{A}^{\prime}$, $x_{B} \neq x_{B}^{\prime}$ and $\mathcal{M}(x_{A}, x_{B}) \neq \mathcal{M}(x_{A}^{\prime}, x_{B}^{\prime})$}
This case is possible only for $M > 2$, because for the case of binary modulation (e.g., BPSK), if $x_{A} \neq x_{A}^{\prime}$ and $x_{B} \neq x_{B}^{\prime}$, both $(x_{A}, x_{B})$ and $(x_{A}^{\prime}, x_{B}^{\prime})$ will lie in the \emph{same} cluster for fixed network coding, i.e., $\mathcal{M}(x_{A}, x_{B}) = \mathcal{M}(x_{A}^{\prime}, x_{B}^{\prime})$ and hence a confusion among these pairs will not cause a symbol error event in the MA phase. For $M > 2$,  $\mathbb{E}(\Upsilon_{1})$ in this case can be written using (7) as
\begin{align}
	& \mathbb{E}(\Upsilon_{1}) = \mathbb{E} \left\{ \exp \left(-\dfrac{E_{s}}{4N_{0}} |h_{A} \Delta x_{A} + h_{B} \Delta x_{B}|^{2}\right)\right\} \nonumber \\
 &  = \int_{0}^{\infty} \int_{0}^{\infty} \int_{-\pi}^{\pi}\exp \left(-\dfrac{E_{s}|\Delta x_{A}h_{A}|^{2}}{4N_{0}} \right) \exp \left(-\dfrac{E_{s}}{4N_{0}} |\Delta x_{B}h_{B}|^{2} \bigg)  \right. \nonumber \\
& \hspace{1cm} \times  \exp \left(-\dfrac{E_{s}\cos \theta |\Delta x_{A} \Delta x_{B} h_{A}h_{B}|}{2N_{0}} \right) f(\theta)f(|h_{A}|) f(|h_{B}|) \ d\theta \ d|h_{A}| \ d|h_{B}|
\end{align}
where $\theta = \angle h_{A} - \angle h_{B}$ is a random variable uniformly distributed over $[-\pi, \pi)$. Using the fact that $\exp(\cos\theta)$ is an even function of $\theta$, integration w.r.t $\theta$ in (20) can be solved as \cite[p.~376]{Stegun}
\begin{align}
	\mathbb{E}(\Upsilon_{1}) & = \int_{0}^{\infty} \exp \left(-\dfrac{E_{s}|\Delta x_{A}h_{A}|^{2}}{4N_{0}} \right)\int_{0}^{\infty} \exp \left(-\dfrac{E_{s}|\Delta x_{B}h_{B}|^{2}}{4N_{0}} \right) \nonumber \\
	& \hspace{4cm} \times \left[ I_{0}\left(\dfrac{E_{s}|\Delta x_{A} \Delta x_{B}|}{2N_{0}}|h_{A} h_{B}|\right)\right] f(|h_{A}|) f(|h_{B}|) d|h_{A}| d|h_{B}| \nonumber 
\end{align}
where $I_{0}(\cdot)$ is the modified Bessel function of the first kind. Substituting for $f(|h_{B}|)$ in the above equation yields 
\begin{align}
	\mathbb{E}(\Upsilon_{1}) & = \int_{0}^{\infty} \exp \left(-\dfrac{E_{s}|\Delta x_{A}h_{A}|^{2}}{4N_{0}} \right) \sum_{l = 1}^{N_{B}} \binom{N_{B}}{l} (-1)^{l-1}  \left[l \int_{0}^{\infty}2|h_{B}| \exp \left\{-\left(l + \dfrac{E_{s}|\Delta x_{B}|^{2}}{4N_{0}}\right) \right. \right.\nonumber \\
	& \hspace{3cm}\times |h_{B}|^{2}\Bigg\} \Bigg] I_{0}\left(\dfrac{E_{s}|\Delta x_{A}\Delta x_{B}|}{2N_{0}} |h_{A}||h_{B}|\right) d|h_{B}| \Big] f(|h_{A}|) d|h_{A}|.
\end{align}

Solving the inner integral using \cite[p.~306]{PrudnikovVol2} yields
\begin{align}
	& \mathbb{E}(\Upsilon_{1}) = \int_{0}^{\infty} \exp \left(-\dfrac{E_{s}|\Delta x_{A}h_{A}|^{2}}{4N_{0}} \right) \sum_{l = 1}^{N_{B}} \binom{N_{B}}{k} (-1)^{k-1}  (\Psi_{B,l})^{-1} \exp \left[\dfrac{\left(\dfrac{E_{s}|\Delta x_{A}||\Delta x_{B}|}{2N_{0}} |h_{A}|\right)^{2}}{4l \Psi_{B,l}}\right] \notag \\
	& \hspace{14cm} \times f(|h_{A}|) d|h_{A}|  \nonumber \\
	& = \sum_{k=1}^{N_{A}} \sum_{l=1}^{N_{B}} \binom{N_{A}}{k} \binom{N_{B}}{l} (-1)^{(k+l-2)}  (\Psi_{B,l})^{-1} k \nonumber \\ 
	& \hspace{3cm}\times  \int_{0}^{\infty} 2|h_{A}| \exp \left[-\left\{\underbrace{\dfrac{E_{s}|\Delta x_{A}|^{2}}{4N_{0}} - \dfrac{\left(\dfrac{E_{s}|\Delta x_{A}||\Delta x_{B}|}{2N_{0}}\right)^{2}}{4l \Psi_{B,l}}}_{\eta} + k \right\} |h_{A}|^{2} \right]  d|h_{A}| \nonumber \\
	& = \sum_{k=1}^{N_{A}} \sum_{l=1}^{N_{B}} \binom{N_{A}}{k} \binom{N_{B}}{l} (-1)^{(k+l-2)} (\Psi_{B,l})^{-1} \dfrac{1}{1+\eta/k} \nonumber \\
	&  = \sum_{k=1}^{N_{A}} \sum_{l=1}^{N_{B}} \binom{N_{A}}{k} \binom{N_{B}}{l} \dfrac{(-1)^{(k+l-2)}}{\left(\Psi_{A,k}\Psi_{B,l} - \dfrac{1}{4kl}\Theta_{A,B}\right)}
\end{align}
where, $\Theta_{A,B} = \left(\dfrac{E_{s}|\Delta x_{A} \Delta x_{B}|}{2N_{0}}\right)^{2}$. Solving in a similar fashion for $\mathbb{E}(\Upsilon_{2})$, we obtain
\begin{align}
	\mathbb{E}(\Upsilon_{2}) = & \sum_{k=1}^{N_{A}} \sum_{l=1}^{N_{B}} \binom{N_{A}}{k} \binom{N_{B}}{l} \dfrac{(-1)^{(k+l-2)}}{\left(\Xi_{A,k}\Xi_{B,l} - \dfrac{1}{4kl}\Phi_{A,B}\right)} 
\end{align}
where, $\Phi_{A,B} = \left(\dfrac{2E_{s}|\Delta x_{A} \Delta x_{B}|}{3N_{0}}\right)^{2}$. Hence, the average SER arising from the case when $x_{A} \neq x_{A}^{\prime}$, $x_{B} \neq x_{B}^{\prime}$ and $\mathcal{M}(x_{A}, x_{B}) \neq \mathcal{M}(x_{A}^{\prime}, x_{B}^{\prime})$ becomes 
\begin{align}
	& P_{e}\{(\hat{x}_{A}, \hat{x}_{B}) = (x_{A}^{\prime}, x_{B}^{\prime}), \mathcal{M}(x_{A}, x_{B}) \neq \mathcal{M}(x_{A}^{\prime}, x_{B}^{\prime})\} \nonumber \\
	& \lesssim \dfrac{1}{M^{2}} \sum_{(x_{A}, x_{B}) \in \mathcal{S}^{2}} \sum_{\substack{x_{A} \neq x_{A}^{\prime} \in \mathcal{S}, x_{B}\neq x_{B}^{\prime} \in \mathcal{S} \\ \mathcal{M}(x_{A}, x_{B}) \neq \mathcal{M}(x_{A}^{\prime}, x_{B}^{\prime})}} \sum_{k=1}^{N_{A}}  \sum_{l=1}^{N_{B}} \binom{N_{A}}{k} \binom{N_{B}}{l} (-1)^{(k+l-2)} \nonumber \\
	& \times \left\{ \dfrac{1}{12}\left(\Psi_{A,k}\Psi_{B,l} - \dfrac{\Theta_{A,B}}{4kl}\right)^{-1}  + \dfrac{1}{4}\left(\Xi_{A,k}\Xi_{B,l} - \dfrac{\Phi_{A,B}}{4kl}\right)^{-1}  \right\} \nonumber \\
	& =  \dfrac{1}{M^{2}} \sum_{(x_{A}, x_{B}) \in \mathcal{S}^{2}} \sum_{\substack{x_{A} \neq x_{A}^{\prime} \in \mathcal{S}, x_{B}\neq x_{B}^{\prime} \in \mathcal{S} \\ \mathcal{M}(x_{A}, x_{B}) \neq \mathcal{M}(x_{A}^{\prime}, x_{B}^{\prime})}}  (\xi_{1} + \xi_{2})
\end{align}
where
\begin{equation}
	\xi_{1} = \sum_{k=1}^{N_{A}}  \sum_{l=1}^{N_{B}} \binom{N_{A}}{k} \binom{N_{B}}{l} \dfrac{(-1)^{(k+l-2)} }{12 \left(\Psi_{A,k}\Psi_{B,l} - \dfrac{\Theta_{A,B}}{4kl}\right)},
\end{equation}
\begin{equation}	
	\xi_{2} = \sum_{k=1}^{N_{A}}  \sum_{l=1}^{N_{B}} \binom{N_{A}}{k} \binom{N_{B}}{l} \dfrac{(-1)^{(k+l-2)} }{4 \left(\Xi_{A,k}\Xi_{B,l} - \dfrac{\Phi_{A,B}}{4kl}\right)}.
\end{equation}

Substituting the values of $\Psi_{A,k}$, $\Psi_{B,l}$ and $\Theta_{A,B}$ into (25), $\xi_{1}$ becomes
\begin{align}
	\xi_{1} = & \sum_{k=1}^{N_{A}}  \sum_{l=1}^{N_{B}} \binom{N_{A}}{k} \binom{N_{B}}{l} \times \dfrac{(-1)^{(k+l-2)}} {12 \left(1 + \dfrac{E_{s}|\Delta x_{A}|^{2}}{4kN_{0}} + \dfrac{E_{s}|\Delta x_{B}|^{2}}{4lN_{0}}\right)} \nonumber \\
	& = \dfrac{1}{12}\sum_{k=1}^{N_{A}}  \sum_{l=1}^{N_{B}} \binom{N_{A}}{k} \binom{N_{B}}{l} \dfrac{(-1)^{(k+l-2)}}{g(k, l) (E_{s}/N_{0})}  \left(1 + \dfrac{1}{g(k, l) (E_{s}/N_{0})}\right)^{-1}
\end{align}
where 
\begin{equation}
	g(k, l) = \dfrac{|\Delta x_{A}|^{2}}{4k} + \dfrac{|\Delta x_{B}|^{2}}{4l}.
\end{equation}

Using the binomial expansion, $\xi_{1}$ can be rewritten as
\begin{align}
	& \xi_{1} = \dfrac{1}{12}\sum_{k=1}^{N_{A}}  \sum_{l=1}^{N_{B}} \binom{N_{A}}{k} \binom{N_{B}}{l} \dfrac{(-1)^{(k+l-2)}}{g(k, l) (E_{s}/N_{0})} \sum_{m = 1}^{\infty} (-1)^{m-1} \left(\dfrac{1}{g(k, l)(E_{s}/N_{0})}\right)^{m - 1} \nonumber  
\end{align}
\begin{align}	
	= & \sum_{m = 1}^{\infty} \left(\dfrac{E_{s}}{N_{0}}\right)^{-m} \underbrace{\dfrac{1}{12}\sum_{k=1}^{N_{A}}  \sum_{l=1}^{N_{B}} \binom{N_{A}}{k} \binom{N_{B}}{l}  (-1)^{k+l+m-3}(g(k, l))^{-m}}_{B_{-m}} \nonumber \\
	= &\sum_{m = 1}^{\infty} B_{-m} \left(\dfrac{E_{s}}{N_{0}}\right)^{-m} < B_{-1} \left(\dfrac{E_{s}}{N_{0}}\right)^{-1} + O \left[\left(\dfrac{E_{s}}{N_{0}}\right)^{-1}\right].
\end{align}

Similarly, for $\xi_{2}$ it can be shown that
\begin{align}
	\xi_{2} < B_{-1}^{\prime} \left(\dfrac{E_{s}}{N_{0}}\right)^{-1} + O \left[\left( \dfrac{E_{s}}{N_{0}}\right)^{-1} \right]
\end{align}
where
\begin{align}
	& B_{-m}^{\prime} = \dfrac{1}{4}\sum_{k=1}^{N_{A}}  \sum_{l=1}^{N_{B}} \binom{N_{A}}{k} \binom{N_{B}}{l}  (-1)^{k+l+m-3}(g^{\prime}(k, l))^{-m}, \\
	& g^{\prime}(k, l) = \dfrac{|\Delta x_{A}|^{2}}{3k} + \dfrac{|\Delta x_{B}|^{2}}{3l}.
\end{align}

From (24), (29) and (30), it is clear that the average symbol error probability due to the case when $x_{A} \neq x_{A}^{\prime}$, $x_{B} \neq x_{B}^{\prime}$ and $\mathcal{M}\left(x_{A}, x_{B}\right) \neq \mathcal{M} \left(x_{A}^{\prime}, x_{B}^{\prime}\right)$ decays as $(E_{s}/N_{0})^{-1}$ for higher values of $E_{s}/N_{0}$. 

The overall average SER for the PNC system with TAS1 given in (5) can be found by adding (12), (18) and (24). It is also clear from (15), (16), (29) and (30) that the first term on the right-hand side of (5) decays as $(E_{s}/N_{0})^{-N_{A}}$, the second term decays as $(E_{s}/N_{0})^{-N_{B}}$ and the third term decays as $(E_{s}/N_{0})^{-1}$. It is important to note that for BPSK case (as in \cite{Huang}), the third term does not contribute to the symbol error at the relay at end of MA phase and hence the overall system diversity order becomes $\min\{N_{A}, N_{B}\}$. For any other modulation with $M>2$, the overall system diversity order for MISO case with TAS1 becomes $\min \{1, N_{A}, N_{B}\} = 1$, resulting in the overall performance degradation. 

\subsection{TAS2: Euclidean distance based antenna selection}
To analyze the performance of TAS2, we follow the approach in \cite{Hari}, where the authors have analyzed the error performance of the ED based antenna selection scheme for SM. The analysis for PNC differs due to that fact that for a SM system, only a single antenna is active during transmission, while for a TWRC with PNC, two antennas (one from each user) transmit simultaneously. 

Given a set of transmit antenna indices $I = (i, j) \in \mathcal{I}$, the set of possible transmit vectors for the PNC system can be defined as $\mathcal{C}_{I} = \{[x_{A} \boldsymbol{e}_{i} \ x_{B} \boldsymbol{e}_{j}]^{T} | x_{A}, x_{B} \in \mathcal{S}\}$, where $\boldsymbol{e}_{i}$ and $\boldsymbol{e}_{j}$ are row vectors of length $N_{A}$ and $N_{B}$ respectively with all zero elements except a 1 at $i^{\text{th}}$ and $j^{\text{th}}$ position respectively. Let $\boldsymbol{z}_{I}(x_{A}, x_{B}) = [x_{A} \boldsymbol{e}_{i} \ x_{B}\boldsymbol{e}_{j}]^{T}$. Defining $\Delta \mathcal{C}_{I} = \Big\{\boldsymbol{z}_{I} \left(x_{A}^{(1)}, x_{B}^{(1)}\right) - \boldsymbol{z}_{I} \left(x_{A}^{(2)}, x_{B}^{(2)}\right) \Big| x_{A}^{(1)}, x_{B}^{(1)}, x_{A}^{(2)}, x_{B}^{(2)} \in \mathcal{S}$, $\mathcal{M}\left(x_{A}^{(1)}, x_{B}^{(1)}\right) \neq \mathcal{M}\left(x_{A}^{(2)}, x_{B}^{(2)}\right) \Big\}$ as the set of difference vectors corresponding to the codebook $\mathcal{C}_{I}$, the set of matrices $\Delta \mathcal{D}$ can be defined as
\begin{equation}
	\Delta \mathcal{D} = \{[\boldsymbol{x}_{1}, \boldsymbol{x}_{2}, \ldots, \boldsymbol{x}_{n}]|\boldsymbol{x}_{k} \in \Delta \mathcal{C}_{k} \forall k \in \{1, 2, \ldots, n\}\}.
\end{equation}
Each element in $\Delta \mathcal{D}$ will be of size $(N_{A} + N_{B}) \times n$ where $n = N_{A} \times N_{B}$. Let $r_{\text{min}}$ be defined as 
\begin{equation}
	r_{\text{min}} = \min \{ \mathrm{rank}(\boldsymbol{X})| \boldsymbol{X} \in \Delta \mathcal{D}\}.
\end{equation}

The minimum number of \emph{linearly independent} columns in $\boldsymbol{X}$, i.e. $r_{\text{min}}$ will be $\min\{N_{A}, N_{B}\}$. To understand this, consider an example when $N_{A} = 3 $ and $N_{B} = 2$. The possible structure of each element in $\Delta \mathcal{D}$ will be 
\begin{equation}
	\boldsymbol{X} = \begin{bmatrix}
	\Delta x_{A}^{(1)} & \Delta x_{A}^{(2)} & 0 & 0 & 0 & 0 \\
	0 & 0 & \Delta x_{A}^{(3)} & \Delta x_{A}^{(4)} & 0 & 0 \\
	0 & 0 & 0 & 0 & \Delta x_{A}^{(5)} & \Delta x_{A}^{(6)} \\ 
	\Delta x_{B}^{(1)} & 0 & \Delta x_{B}^{(3)} & 0 & \Delta x_{B}^{(5)} & 0\\
	0 & \Delta x_{B}^{(2)} & 0 & \Delta x_{B}^{(4)} & 0 & \Delta x_{B}^{(6)}
	\end{bmatrix}.
\end{equation}

Thanks to the definition of $\Delta \mathcal{C}_{I}$, in which the condition $\mathcal{M}\left(x_{A}^{(1)}, x_{B}^{(1)}\right) \neq \mathcal{M}\left(x_{A}^{(2)}, x_{B}^{(2)}\right)$ ensures that $\boldsymbol{z}_{I} \left(x_{A}^{(1)}, x_{B}^{(1)}\right) \neq \boldsymbol{z}_{I} \left(x_{A}^{(2)}, x_{B}^{(2)}\right)$ and hence in any of the column of $\boldsymbol{X}$, both $\Delta x_{A}^{(i)}$ and $\Delta x_{B}^{(i)}$ cannot be zero simultaneously. If $\Delta x_{B}^{(1)}$ and $\Delta x_{B}^{(2)}$ are non-zero, they form a non-zero minor (a diagonal matrix) and the minimum possible rank of $\boldsymbol{X}$ becomes 2. Now if $\Delta x_{B}^{(1)}$ is zero and any one from $\Delta x_{B}^{(3)}$ or $\Delta x_{B}^{(5)}$ is non-zero then also a $2 \times 2$ non-zero minor can be formed using $\Delta x_{B}^{(2)}$. A similar argument applies when $\Delta x_{B}^{(1)}$ is non-zero and $\Delta x_{B}^{(2)}$ is zero and a $2 \times 2$ minor can be formed using $\Delta x_{B}^{(1)}$ and $\Delta x_{B}^{(4)}$ or $\Delta x_{B}^{(6)}$ with the help of a column swap. In the case $\Delta x_{B}^{(1)}$, $\Delta x_{B}^{(3)}$ and $\Delta x_{B}^{(5)}$ are zero, the $\Delta x_{A}^{(i)}$ values in the corresponding columns will be non-zero and they will form three linearly independent columns and the rank of matrix $\boldsymbol{X}$ will be at least 3. A similar argument is valid for the case when $\Delta x_{B}^{(2)}$, $\Delta x_{B}^{(4)}$ and $\Delta x_{B}^{(6)}$ are zero. On the other hand, if all the $\Delta x_{B}^{(i)}$ values are zero then the first three rows will be linearly independent and the rank of $\boldsymbol{X}$ will be 3. Hence the minimum possible rank of $\boldsymbol{X}$ is 2 i.e., $\min \{N_{A}, N_{B} \}$. It is straightforward to generalize this argument to the case of an arbitrary number of antennas at each user.

Let the transmit vectors in the each codebook be denoted as $\mathcal{C}_{k} = \{\boldsymbol{x}_{l}(k)|l \in \{1, 2, ,\ldots, M^{2}\}\}$ and the optimal set of transmit antennas for any particular channel realization $\boldsymbol{H}$ be $I_{k^{\ast}}$. For $E_{s}/N_{0} \gg 1$, the average pairwise error probability between any two different transmit vectors indexed by $l_{1}$ and $l_{2}$ in the codebook $\mathcal{C}_{k^{\ast}}$ can be expressed, using the Chernoff bound, as \cite[eqn.~(4)-(10)]{Hari}
\begin{align}
	\text{PEP}(\boldsymbol{x}_{l_{1}} \rightarrow \boldsymbol{x}_{l_{2}}) \leq & \dfrac{1}{2} \left(\dfrac{E_{s}\lambda^{\ast}}{4nN_{0}} \right)^{-r_{\text{min}}}.
\end{align}
where $\lambda^{\ast} = \min_{\boldsymbol{X} \in \Delta \mathcal{D}} \lambda_{s}(\boldsymbol{XX}^{H})$ and $\lambda_{s}(\boldsymbol{Y})$ denotes the smallest non-zero eigenvalue of matrix $\boldsymbol{Y}$. An upper bound on the average SER for TAS2 at $E_{s}/N_{0} \gg 1$ can therefore be given as
\begin{align}
	P_{e} \leq & \ \dfrac{1}{2M^{2}} \sum_{\boldsymbol{x}_{l_{1}} \in \mathcal{C}_{k^\ast}} \sum_{\substack{\boldsymbol{x}_{l_{1}} \neq \boldsymbol{x}_{l_{2}} \in \mathcal{C}_{k^{\ast}} \\ \mathcal{M} \left(x_{A}^{(l_{1})}, x_{B}^{(l_{1})}\right) \neq \mathcal{M} \left(x_{A}^{(l_{2})}, x_{B}^{(l_{2})}\right)}}  \left(\dfrac{E_{s}\lambda^{\ast}}{4nN_{0}} \right)^{-r_{\text{min}}} \nonumber	 \\
		= & \ \binom{M}{2} \left(\dfrac{E_{s}\lambda^{\ast}}{4nN_{0}} \right)^{-\min \{N_{A}, N_{B}\}}.
\end{align}

It is clear from (37) that the PNC system with TAS2 achieves a diversity order lower bounded by $\min(N_{A},N_{B})$ for any modulation order $M$.

\section{Results and Discussion}
In this section, we present a performance comparison of the two TAS schemes discussed in the previous sections. We consider the scenario where both users transmit to the relay using a unit-energy QPSK constellation. In Fig. 4, the SER performance for the two TAS schemes is shown with different number of antennas at the users. It is clear from the figure that for TAS1, the system achieves a diversity order equal to 1 irrespective of number of antennas, as for the higher value of $E_{s}/N_{0}$ the SER curve becomes parallel to $(E_{s}/N_{0})^{-1}$ in each case. It is also worth noting that the derived closed-form expression for the upper bound on the SER is very tight for TAS1. 
\begin{figure}[hbht]
\centering
\includegraphics[scale=0.9]{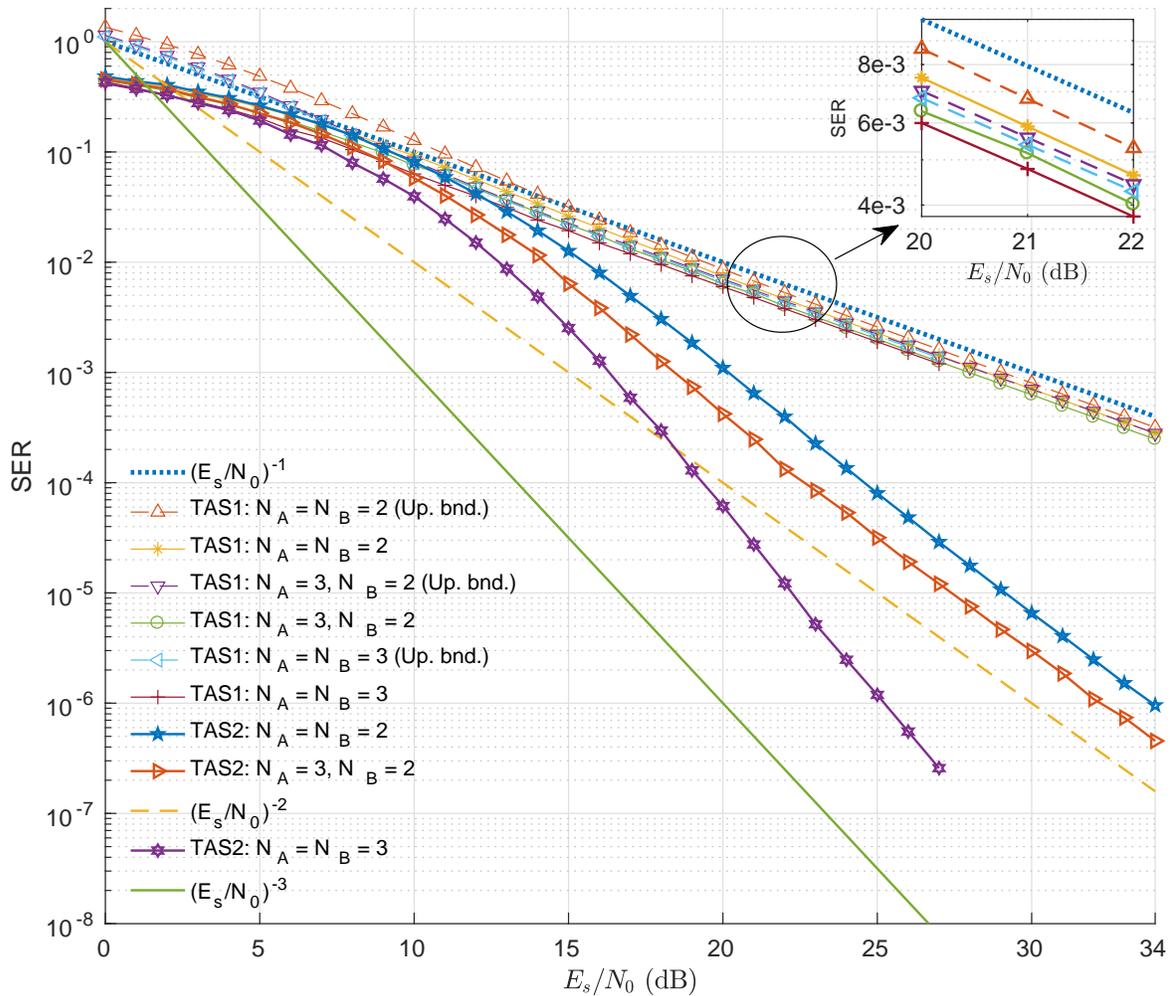}
\caption{Performance comparison of symbol error rate for the two TAS schemes.}
\end{figure}

In contrast to this, the PNC system with TAS2 outperforms the one with TAS1 while achieving a higher diversity order. For the case when $N_{A} = N_{B} = 2$ and $N_{A} = 3$, $N_{B} = 2$, the average SER in the PNC system with TAS2 decays more rapidly as compared to TAS1 and becomes parallel to $(E_{s}/N_{0})^{-2}$ for higher values of $E_{s}/N_{0}$. Similarly, for the case when $N_{A} = N_{B} = 3$ the average SER for TAS1 decays with $(E_{s}/N_{0})^{-1}$ while for TAS2 the average SER decays with $(E_{s}/N_{0})^{-3}$ at higher values of $E_{s}/N_{0}$. 

\section{Conclusion}
In this paper, we provided a comprehensive error performance analysis for a fixed network coded MISO PNC system under two different TAS schemes. It was shown that for the first TAS scheme, where antenna selection is performed based on the strongest channel, the system performance is severely degraded for $M > 2$ due to the existence of singular fade states, and the overall system diversity order remains equal to one irrespective of the number of antennas at the users. A closed-form expression for a tight upper bound on the average SER was also presented. Furthermore, we propose a new TAS scheme in which the antenna selection is performed such that it maximizes the minimum ED between the clusters. The PNC system using this new antenna selection scheme efficiently mitigates the deleterious effect of singular fade states without any need of adaptive network codes or unconventional constellation design, reducing the overall system complexity significantly. We also show that the diversity order of such a PNC system at the end of the MA phase is lower bounded by the minimum of the number of antennas at the two users.
\section*{Acknowledgment}
This publication has emanated from research conducted with the financial support of Science Foundation Ireland (SFI) and is co-funded under the European Regional Development Fund under Grant Number 13/RC/2077.

\bibliographystyle{IEEEtran}
\bibliography{OneColumn}
\end{document}